\documentclass[a4paper,12pt]{article}
\usepackage{amssymb,amsmath,mathbbol,mathrsfs}
\usepackage[usenames,dvipsnames]{color}
\usepackage{hyperref}
\usepackage{xcolor}
\usepackage{stmaryrd}
\usepackage{authblk}
\usepackage{framed}
\usepackage{empheq} 
\usepackage{slashed}
\usepackage{hyphenat}
\usepackage{cite}
\usepackage{apacite}
\usepackage{natbib}
\usepackage{dsfont}
\usepackage{chngcntr}
\usepackage{enumerate}  

\usepackage{marginnote}    

\usepackage[left=.8in,right=.8in,top=.8in,bottom=.8in]{geometry}                  % For good copy.

\usepackage{sectsty} 
\allsectionsfont{\sffamily\mdseries\upshape} % (See the fntguide.pdf for font help)
\usepackage{tocloft}

\makeatletter
\renewenvironment{abstract}{%
    \if@twocolumn
      \section*{\abstractname}%
    \else %% <- here I've removed \small
      \begin{center}%
        {\bfseries\sffamily\abstractname\vspace{\z@}}%  %% <- here I've added \Large
      \end{center}%
      \quotation
    \fi}
    {\if@twocolumn\else\endquotation\fi}
\makeatother

\numberwithin{equation}{section}
\hypersetup{
	colorlinks=true,         
	linkcolor=MidnightBlue,          
	citecolor=BrickRed,        
	urlcolor=MidnightBlue            
}

\newcommand{\be}{\begin{equation}}
\newcommand{\ee}{\end{equation}}

\newcommand{\F}{{{\Phi}}}
\renewcommand{\d}{{\mathrm{d}}}

\newcommand{\G}{{\mathcal{G}}}

\newcommand{\pp}{{\partial}}

\renewcommand{\bar}{\overline}

\newcommand{\RR}{\mathds{R}} %Isham-style

\newtheorem{defi}{Definition}

\newcommand{\cint}{{\int\kern-.87em{<}}}
\newcommand{\sint}{{\int\kern-.75em{\sim}}}
\newcommand{\fint}{{\int\kern-1.00em{\int}}}

\newcommand{\bb}{\mathbb}

\let\oldmarginpar\marginpar
\renewcommand\marginpar[1]{\oldmarginpar{\color{red}\raggedright\footnotesize #1}}

\title{  {\huge Same-diff?}\\  Conceptual similarities between gauge transformations and diffeomorphisms\\
{\large Part II: Challenges to sophistication}
}
\author{Henrique de A. Gomes\footnote{\href{mailto:gomes.ha@gmail.com}{gomes.ha@gmail.com}} \\\it University of Oxford\\ \it Oxford, OX2 6HT, UK
}

\begin{document}

\maketitle
\begin{abstract}
  %The famous hole argument poses a problem of under-determination in a theory embodying diffeomorphisms as symmetry transformations. The most popular resolution of this problem is \textit{sophisticated substantivalism}, also called anti-haecceitism: a form of structuralism, taking spacetime points to have no metaphysically robust identity across possibilities; points may acquire identity only through the complex web of inter-relations of objects (including fields) in spacetime. But this resolution is much less popular when it comes to the under-determination arising from gauge symmetry. For gauge symmetry, attempting to \textit{eliminate} the symmetry-related models is also a popular option. 
The following questions are germane to our understanding of gauge-(in)variant quantities and  physical possibility: how are gauge transformations and spacetime diffeomorphisms understood as symmetries, in which ways are they similar, and in which are they different? To what extent are we justified in endorsing different attitudes---nowadays called sophistication, haecceitism, and eliminativism---towards each?   This is the second of four papers taking up this question, and it is the one that most engages with the metaphysical debates surrounding our understanding of symmetry and equivalence. 

 In this paper, I will provide two desiderata for the application of a treatment of symmetries known as `sophistication' and show that both general relativity and Yang-Mills theory satisfy these desiderata. The first desideratum for symmetries is mathematical, and was shown to hold for general relativity and Yang-Mills in the first paper \citep{Samediff_0}: (i) that they correspond to the automorphisms of the structured base sets for the models of the two theories.  Here I will extend the desideratum to more general theories. The second desideratum  is mathematical-physical: (ii) that the general type of structure to which Desideratum (i) refers is axiomatizable and that this axiomatization can be phrased in terms of basic  physical predicates of the theory. In the third paper in the series, \cite{Samediff_1b}, I will provide yet a third desideratum, to deal with an issue that goes beyond the standard debate about sophistication. %In the fourth and last paper of the series \cite{Samediff_2}, I proceed in more technical, low-altitude detail, comparing the symmetries of general relativity and Yang-Mills theory.   %  Items (iii) and (iv) require a more operational interpretation of symmetries---very similar to Einstein's original understanding of diffeomorphisms/coordinate transformations---which relates the  active and passive transformations. 
 %Fulfillment of these desiderata require a `relational' understanding of coordinates (or frames, etc): a correspondence between the isomorphisms of the theory and choices of which qualitative relations are used to parametrize  physical  observables. Thus they allow a concise description of structure-tokens, not just structure-types. 
 
   %While the extant literature has argued that the two kinds of symmetry discussed here share many features,  it has mainly focused on (i), and on a metaphysical thesis of \emph{anti-haecceitism}, or \emph{anti-quidditism}, which I take to be most naturally enforced by the entire collection (i)-(iv). 
 
  %There is no clear path from this classification towards a more strict attitude---such as reduction---towards gauge.  Considerations of other explanatory and theoretical virtues---such as explicit Lorentz covariance or compositionality---militate for sophistication about, not elimination of, both diffeomorphisms and gauge transformations. Thus I advertise  a structuralist interpretation of the connection-form on a principal fiber bundle, on par with a structuralist interpretation of the metric on spacetime; that is, I advertise anti-quidditism on a par with anti-haecceitism. 
 
\end{abstract}

{\center \begin{quote}Same-diff [noun]: \textit{
an oxymoron, used to describe something as being the same as something else. Often used as an excuse for being wrong.} (Urban dictionary). 

%Diff: \textit{A common abbreviation for ``diffeomorphism''. E.g. Diff$(M)$ is the group of diffeomorphisms of the (differentiable) manifold $M$.} 
\end{quote}}
\tableofcontents
\bigskip

\newpage

\section{Introduction}\label{sec:intro}

This is the second of three papers analysing the similarities and distinctions between the gauge symmetries of Yang-Mills theory and the spacetime diffeomorphisms of general relativity. The first three  analyse more formal aspects while the fourth will analyse more detailed aspects  of this comparison.  The previous paper, \citep{Samediff_0}, has given general definitions of dynamical symmetries and applied them to the specific cases of Yang-Mills and general relativity. In this paper I will try to provide more general criteria for the interpretation of symmetries, with a more metaphysical focus and taking these two theories as templates.

\subsection{Motivation}\label{sec:mot}

Gauge theories lie at the heart of modern physics: in particular, they constitute the standard model of particle physics.
Philosophers of physics generally accept as the leading idea of a gauge theory---or as the main connotation of the phrase `gauge theory’---that it involves a formalism that uses more variables than there are physical degrees of freedom in the system described; and thereby more variables that one strictly speaking needs to use. Hence the common soubriquets: `descriptive redundancy’, `surplus structure’, and more controversially, `descriptive fluff’ (e.g. \cite{Earman_gmatters, Earman2004}). 

Although the main idea and connotation of descriptive redundancy has been endorsed by countless presentations in the physics literature, some celebrated philosophers, such as \citet{Healey_book} and \citet{Earman_gmatters} among others, have gone beyond this connotation, and defended a stronger,  \emph{eliminativist view}. The view is that  gauge symmetry must be  `eliminated' before determining which models of a theory represent distinct physical possibilities, on pain of radical indeterminism. For them,  the connotation of `fluff' is that it can have no purpose. 

But radical indeterminism also threatens theories such as general relativity, embodying diffeomorphism symmetry.  There the threat---which underlies the infamous `hole argument' \citep{EarmanNorton1987}---finds its most  popular resolution in a treatment of, or rather, an attitude towards, symmetry-related models, called \textit{sophisticated substantivalism}. 

 Sophisticated substantivalism  is \emph{not} eliminativist; but it is \emph{anti-haecceitist}. This jargon can be quickly summarized: haecceitism is the doctrine that objects have an intrinsic identity (or `thisness': {\em haecceitas}); haecceitistic possibilities involve individuals being ``swapped'' or ``exchanged'' without any qualitative difference. Similarly,  \emph{quidditistic} possibilities involve properties  being ``swapped'' or ``exchanged'' without any qualitative difference. Anti-haecceitists about
spacetime points thus deny that there are possible  worlds that instantiate the same distribution of qualitative
properties and relations over spacetime points, yet differ only over which
spacetime points play which qualitative roles. Similarly, the anti-quidditist will insist that there are no two possible worlds
 that instantiate the same nomological structure, and yet differ only over which properties play which nomological roles.\footnote{Caspar Jacobs points out to me that there may here be some tension between the standard use of qualitative and quidditism. for a couple of general philosophy topics this term. The properties that are swapped under quidditism will, elsewhere, be classified as qualitative (e.g. mass values). \cite{Jacobs_guide} proposes, as a solution, that anti-haecceitism, understood as the claim that there are no distinct yet qualitatively identical possibilities, suffices. I agree, but for the sake of the dialectic between gauge and gravity, I will maintain the terminology of `quiddistic',  for those possibilities that differ only over which properties play which nomological roles.}%  \citep{Black_quidditism} is a standard example of the anti-quidditist position, while \citep{LewisRamsey} is an example of the quidditist one.

 In sum, according to the doctrine of sophisticated substantivalism,  possible worlds can only be distinguished by qualitative properties and spacetime points  have no metaphysically robust identity across possibilities---points can \textit{only} be singled out or individuated through their complex web of properties and  relations, as encoded in fields.\footnote{It is not a given that points \emph{can be} singled out (equivalently: uniquely specified or individuated)  by such  a web of properties and relations:  if this web is not sufficiently complex, the specification will fail. This type of obstruction to individuation due to  homogeneity  is well-known, and related to \cite{Black_PII}'s criticism of Leibniz's Principle of the Identity of Indiscernibles. See \citep[Sec. 3.3]{Pooley_draft}  for a thorough exposition, and \cite{Wuthrich_abysmal} for a related debate in the context of general relativity (we will come back to this topic in \cite{Samediff_1b}). \label{ftnt:complex}} 

A similar resolution is available for gauge symmetry, in the form of `anti-quidditism'; but it is  much less popular in that context.\footnote{But recently the position has garnered support, starting with \cite{Dewar2017} and followed by \cite{ReadMartens, Jacobs_thesis, Jacobs_Inv}.} In the case of gauge symmetry, attempting to \textit{eliminate} the symmetry-related models is considered a more viable alternative. But is this alternative really more justified in the case of gauge symmetry? If so, why? \\

\subsection{Morals of this paper}\label{sec:method}

Given a theory with   local  gauge symmetry, the  eliminativist seeks a second theory,  empirically equivalent to the first but with possibly different ontological commitments, that has no local gauge symmetry.
But not all is lost if   eliminativism fails:  the threat of a pernicious type of indeterminism can still be countered by \emph{sophistication}. This is the position that, generalizing sophisticated substantivalism, accepts that there are isomorphic models within a theory's formalism but simply  denies that these models represent distinct physical possibilities.
And, in both the older and the more recent philosophical literature about isomorphisms in general relativity,  sophistication is, in effect  if not in name,  reported to be the position of the majority of  theoretical physicists.\footnote{As reported by e.g.  \cite[p. 438]{Wald_book}, \cite[p. 5]{Oneill}, \cite[p. 68]{HawkingEllis}. 
The literature about the treatment of isomorphic models and indeterminism in general relativity is vast, and, since the 1980s, tied to the  `hole argument' (see \cite{EarmanNorton1987, Butterfield_hole, Brighouse_hole,  Hoefer_hole} for discussions). More recently, the topic has been  revived by \cite{Weatherall_hole} and several responses to it: \cite{Fletcher_hole, Roberts_hole, Curiel2018, Pooley_Read, GomesButterfield_hole}; see \cite{RobertsWeatherall_hole} for a collection of responses.}  %it has been claimed that, ever since Einstein's own discussion of the hole argument (see \cite{Janssen, EarmanNorton1987} for a description),

 But there are (at least) three  worries about sophistication that so far have not been adequately addressed in the literature. 
 
The first two indict sophistication as being only a half-way house, as too abstract to shed light on  the underlying ontology of the theory. Within this theme,  the first main worry---see e.g. \cite{Moller, Dewar2017, ReadMoller, Jacobs_Inv}---is about whether sophistication  can be obtained all too easily, by just stipulating that symmetry-related models of any theory represent the same physical possibility, even without a clear understanding of the symmetry-invariant ontology. This  worry is that this stipulation still leaves  the inference from  mathematical  isomorphism to  physical (or metaphysical) equivalence   opaque or even unjustified. For instance, in the familiar case of vacuum general relativity, it questions whether we are right in assigning physical, chronogeometric significance to diffeomorphism-invariant quantities.

 I will call this worry, which finds echoes in the literature on the metaphysics of spacetime \citep{Dasgupta_bare, Teitel_hole},  Worry (1), or  `the obscurity of structure-type'.  It besets the philosopher more than the physicist, and it will be the focus of this paper.\footnote{This worry is the same as \cite{Moller, ReadMartens} objection to interpretationalism and thus favoring `motivationalism', as we will see in Section \ref{par:cheap}.} 
 
 But this first worry does not question whether or not we have perspicuous representations of each such equivalence class. For, even if we agree on the given notion of isomorphism, and agree on what kind of things are invariant under this notion, we may still not be able to characterize physical possibilities using just those things.  For instance, again in the familiar case of vacuum general relativity, even if we agree to assign physical, chronogeometric significance to diffeomorphism-invariant quantities, we still may not know how to specify a particular physical possibility using just diffeomorphism-invariant predicates. In other words, a resolution of the first worry does not provide a set of quantities whose values label the isomorphism classes. The worry that, indeed, we cannot do so is the second worry: which besets the  physicist more than the philosopher.

The second worry---about the obscurity of structure-tokens, or Worry (2)---is closer to the physicist's heart, because, as described in \cite{belot_earman_1999, belot_earman_2001}, when theoretical physicists---especially those working in quantum gravity---get down to brass tacks, they need to `get inside' each world; they seek more perspicuous,  piece-wise characterizations of the symmetry-invariant ontology, even while endorsing sophistication at a more general level.
 This worry, and its resolution, will be the focus of  the third paper, \cite{Samediff_1b}. 

The third worry should be concerning to both the philosopher and the physicist. As \cite{Belot50} forcefully argues, in certain sectors of general relativity and Yang-Mills theory, and contrary to sophistication, certain isomorphisms \emph{are}  taken to relate different physical possibilities. We will also sketch a reply in \cite{Samediff_1b} (see also \cite{DES_gf}). A more thorough reply will be deferred to forthcoming work, since it requires a detailed analysis of subsystems in general relativity. 

%So theoretical physicists do not form a single monolithic block, and some, in some circumstances, would question the familiar or standard view of physical equivalence between isomorphic models---labeled \emph{Leibniz equivalence} in the philosophical literature (cf. \cite{EarmanNorton1987}). Another distinction amongst physicists is between those who work at a more abstract and those who work at a more concrete  level.  And the view of the ones that work at the concrete level also runs up against the conceptual understanding of the majority of philosophers of physics: contrary to the picture of isomorphisms relating distinct models of the universe many practicing physicists tend to see only a mathematically harmless---but physically salient---redundancy of choices of coordinates with which we describe physical systems. That is, these physicists, much like Einstein himself (see e.g. \cite{Norton_EinsteinCov}), tend to construe isomorphisms `passively', as saying that  different material coordinate systems can describe the same underlying physics. Whereas philosophers (and the more abstract-minded physicist) tend to construe the symmetries `actively': as providing an isomorphic mathematical model of the same spacetime.

My hope is that, jointly, the first three of the four papers comparing gauge symmetries and diffeomorphisms will  further  illuminate the doctrine of sophistication for both diffeomorphisms and the gauge symmetries of Yang-Mills theory. % \cite{Samediff_1a} will  through an `operational' lens, thus making the doctrine less obscure to the concrete-minded physicist and thereby unifying different understandings of symmetry. This lens provides
 This will be attained by providing further desiderata for sophistication in general (namely, in the list below), beyond those that have been  clearly articulated in the literature---that are mostly included in Desideratum (i) in the list below.

 The desiderata can be divided into the purely mathematical (i), and the  physico-mathematical  (ii and iii). 
\begin{enumerate}[(i)]
\item That the symmetries be induced by the automorphisms of some `natural' geometric structure; 
\item That this geometric structure is axiomatizable, and that the axiomatization can be phrased in terms of basic physical predicates assumed for theory.
\item That the automorphisms of the structure correspond to changes between different choices  of physical coordinate systems, or physical reference frames.
\end{enumerate}
Desideratum (i) ensures the invariant structure of the models admits a natural `qualitative' interpretation.  Desideratum (ii) ensures the invariant structure is `metaphysically perspicuous': a property that has been left rather vague within the literature (cf. Section \ref{par:cheap}).  

Both general relativity and Yang-Mills theory (and also Newton-Cartan theory, which will not be treated here) satisfy all three desiderata.\footnote{ And indeed, in these cases, satisfaction of (iii) follows from the satisfaction of (i). The underlying reason is the proof, contained in \cite[Sec. 5]{Samediff_0}, of a 1-1 correspondence between  dynamical symmetries (or a certain notion thereof: see Definition 2 in \cite[Sec. 2]{Samediff_0}) and merely notational changes, such as changes of bases of a vector space or changes of local coordinate systems. But I am open to the possibility that the method of proof is limited, and there may be theories for which this correspondence does not hold. Thus I distinguish the two desiderata.} 

 This second paper will discuss (i) and (ii), but not (iii), which is aimed at the second worry (about structure-tokens) and which is a formalization of Einstein's original understanding of coordinate systems (cf. \cite{Norton_EinsteinCov}). I will leave a discussion of (iii) for the third paper, \cite{Samediff_1b}.

  The first three papers in this series should be interesting even for those who are only concerned with the interpretation of isomorphisms in general relativity and `the hole argument', and have little interest in gauge theory; it is only in  \cite{Samediff_2} that I will focus on more detailed contrasts between the symmetries of the two theories. \\

 %Pursuing this aim will also close an interpretational gap between the isomorphisms of Yang-Mills theory and general relativity. 

 Here is a brief outline about how I plan to proceed.  In Section \ref{sec:soph_cheap} we challenge the sophisticationist doctrine, with several requests for clarification. In Section \ref{sec:responses}, we  respond to these requests by providing sharpened desiderata for sophistication and showing that both Yang-Mills theory and general relativity meet them. In %Section \ref{subsec:skeptic_why2} I gather threads, and expound what the resolution of the obscurity of structure type  leaves unresolved; and in
  Section \ref{sec:conclusions_soph} I conclude.

\section{Formal desiderata for sophistication: two challenges }\label{sec:soph_cheap}

%In Section \ref{subsec:soph_anti} we briefly exhibited some `metaphysically perspicuous'  interpretations of the structure of models of general relativity, that were taken to render sophisticated substantivalism  plausible. Here we will expand on and challenge those arguments. 
In this paper I  investigate the relation between symmetry-related models and physical possibilities. The general attitude towards this relation that I will defend, introduced briefly in Section \ref{sec:intro}, is what I, and the literature, call `sophistication'. My main aim in this paper is to  find good desiderata for sophistication, in general, and not just for the familiar cases of general relativity and Yang-Mills (as well as other theories, e.g. Newton-Cartan theory).  But if we want to apply the desiderata in general, we must first verify that they apply to these particular cases that have been well-studied in the literature. 

 Having described the symmetry structure of both Yang-Mills theory and general relativity in \cite[Secs. 3, 4]{Samediff_0},  and verified that they meet the formal definitions of symmetry outlined in \cite[Sec.2]{Samediff_0}, and having there  also developed a plausible structural interpretation of both theories, I will now develop a more general  formalization of sophistication, and check whether it is also clarifying in the familiar cases.  
 
First, in Section \ref{sec:antihaec}, expanding the initial remarks of Section \ref{sec:intro}, I  describe the doctrine of sophistication and anti-haecceitism in more detail, and pose some initial concerns. %discuss more formal desiderata for a mathematical structure to support a sophisticated interpretation.% But I should warn the reader that this Section is more speculative, and  also more metaphysical, than most of the paper. 

 In order to better grasp the validity of sophistication in general, I will expand on the first worry about extending  sophistication from general relativity to other theories: the worry about structure-type (Worry (1)). This worry  is composed of two parts, which will be dealt with in this paper. Broadly:\\
\noindent Worry (1-i):  is the isomorphism-invariant structure of the theory sufficiently clear?\\
\noindent Worry (1-ii): what warrant does the isomorphism-invariant part of the theory have  to be the only referring part of the theory? \\
In the extant literature, the first worry is related to the debate about `internal and external sophistication';  and the second worry is often read as a request for a  `metaphysically perspicuous' interpretation of the symmetry-related models. I will expound these worries in  Section  \ref{sec:soph_cheap2}.%---namely, that their core ontology be based on a definition of a symmetry-invariant structure---

%I then argue  that a straightforward formal criterion for ais either  too strict or too abstract to fully convince the skeptic about structuralism. %The skeptic I have in mind here will thus challenge the sophisticationist to clearly articulate what the invariant structure is, for any theory.
  Then, in Section \ref{sec:responses} I will turn to answering the skeptic, by propounding my two Desiderata (i) and (ii), that answer (1-i) and (1-ii). Though I could identify reflections of Desideratum (i) in several places in the literature on sophistication, I believe Desideratum (ii) has not yet made an appearance, in any guise. %In brief, Desideratum (i) guarantees that the symmetries of the theory coincide with isomorphisms of the models of the theory, and these isomorphisms are induced by the automorphisms of the base set of the models of the theory (which is usually a natural geometric structure). Desideratum (ii) guarantees that the structure that is invariant under the isomorphisms can be defined axiomatically, where   the axioms  must be formulated in terms of  basic physical posits of the theory.  
  
   %By thus  reassessing the case of general relativity, {\old I then present a resolution in terms of a correspondence between passive and active symmetry transformations, in Section  \ref{sec:gr_re}. In Section  \ref{sec:soph_antiq} I apply this resolution to the case of Yang-Mills theory, and provide the corresponding metaphysically perspicuous interpretation of the theory.}

  %Now, if $\gamma$ is a curve that interpolates between points $p$ and $q$, and has a certain length according to a metric, under a dragging by $f\in\Diff(M)$, the image of $\gamma$, that interpolates between the image of the points $p$ and $q$, will have the same length according to the dragged metric.
 \subsection{Anti-haecceitism about spacetime in more detail}\label{sec:antihaec}\label{subsec:soph_anti}

How should we, within a given dynamical theory,   interpret models that are related by a dynamical symmetry? The question  has a respectable ancestry: it descends from the centuries-old, absolute-relational
debate, about the nature of space, that began with the dispute between Newton and Leibniz. In the  correspondence between  Clarke---Newton's spokesman---and Leibniz, the question of whether there are alternative possible universes that are qualitatively identical but have their material content translated in space was a major point of contention. According to anti-haecceitism,   since these two models for the universe are qualitatively identical and, in some sense, isomorphic,   they  represent the same physical possibility.

Let us remind the reader of our brief description of  haecceitism in Section \ref{sec:mot}: it  is the doctrine that possibilities can match qualitatively but differ merely over \emph{which individual} has which set of qualitative properties. Thus anti-haecceitists about
spacetime points  deny that there are possible  worlds that instantiate the same distribution of qualitative
properties and relations over spacetime points, yet differ only over which
spacetime points play which qualitative roles. The notion of qualitative is itself contentious and used rather as a term of art in metaphysics (cf. \cite{Adams1979}). But as a ball-park approximation, we can verbally grasp non-qualitative distinctions by the use of 
singular propositions: these distinctions typically require proper names, indexicals or other such devices for their expression. We will regiment the idea of `qualitative' property for certain physical theories in Definition \ref{def:qualitative}, in  Section \ref{sec:responses} (see also footnote \ref{ftnt:qualitative}).\footnote{In this series of papers,  I mean `properties' in the more general sense: not solely as unary, or 1-place relations, but general n-place relations. This was traditionally dubbed an `attribute', but I think this term is ever so slightly more abstract than  `property', and has fallen out of use; and so I will stick with `property'.  }
%Such singular terms require reference-fixing definitions that are  not   functions of the properties and relations of the referent.  

 For now, to gather some intuition, let us briefly look at the corresponding debate about Newtonian absolute space. There, anti-haecceitism would rule  as identical two universes whose material content was uniformly translated, for the theory has no preferred origin of Newtonian space. That is,  space in this theory is modeled by the Euclidean affine space $\mathbb{E}^3$, which contains translations as some of its isomorphisms. So, if we take one model's matter distribution and  obtain another by translating that distribution five feet to the left, these two distributions will have the same relation to the ambient, translation-invariant, Euclidean geometry. Therefore, any difference between the two models must be expressed using indexicals, or other singular terms, e.g. that fix the reference of what `left' is and where the origin of $\mathbb{E}^3$ is in each model. But under any such comparison, a qualitative predicate for a particle, such as `being at the center of mass', retains its truth value under a uniform  translation of all particle positions.\footnote{  There are important differences between positions and velocities: Newtonian space comes equipped with \emph{an absolute origin for velocities}---`being at rest' is a qualitative fact. Therefore anti-haecceitism will identify universes related by a static shift---i.e. a time-independent spatial translation, as described above---but not by a kinematic shift, such as a Galilean boost; see \cite{MaudlinBuckets} and \cite{sep-newton-stm} for more about this distinction. \label{ftnt:abs_vel}}  
%Similarly, 

One consequence of the anti-haecceitist view is that spatial points have no \emph{primitive} identity across models or worlds; such identity would necessarily be non-qualitative. 
The position labeled \emph{sophisticated substantivalism}, introduced in the context of the hole argument in general relativity \citep{EarmanNorton1987},    is anti-haecceitist in this way. %, and illustrates   structuralism for general relativity (but see footnote \ref{ftnt:Ladyman}), since it awards physical significance only to any quantity that is   diffeomorphism-invariant.
What the \emph{sophisticated substantivalist} adds to anti-haecceitism is a commitment to a  spacetime manifold, as it is usually understood. That is,  the view takes the denial of primitive identity for spacetime points to be  perfectly compatible with the existence of a spacetime manifold, composed of spacetime points. 

In the words of \citet[p. 20]{Hoefer_hole}:
 “[p]rimitive identity is metaphysically
otiose, and not a necessary part of the concept of a substance”. In other words,  we should not imbue  every aspect of our set-theoretic constructions with  referential significance. 
 \citet[p. 96]{Kaplan1979} (as quoted in \cite[p. 64]{Pooley_draft}) explains why it may be psychologically appealing  to adopt a metaphysics in which points have primitive identity irrespective of their properties and relations, i.e. irrespective of their ``clothing'': \begin{quote}
``[T]he use of models as representatives of possible worlds has become so natural for logicians
that they sometimes take seriously what are really only artifacts of the model. In particular,
they are led almost unconsciously to adopt a bare particular metaphysics. Why? Because the
model so nicely separates the bare particular from its clothing. The elements of the universe of
discourse of a model have an existence which is quite independent of whatever properties the
model happens to tack onto them.  \end{quote} 
I find Kaplan's explanation, and warning, convincing. That is, \emph{pace} the temptations from model theory,  there is no clear-cut separation between the points and their clothing. Nonetheless, the sophisticated substantivalist would say, the clothed points of spacetime exist!
 
% More recently,  \citet[p.959]{Belot50} labelled this type of position: ``straight-up cheap anti-haecceitism", which he characterized as adopting ``a qualitative counterpart theory and [...denying] the existence of worlds that are qualitative duplicates of one another. Then one can maintain that there is only one possible world corresponding to a whole family of mathematical spacetimes related to one another by generalized shifts. The threat of indeterminism vanishes [...]"
% In other words, anti-haecceitism discards any primitive identity of spacetime points across models or worlds; the sophisticated substantivalist argues this  position still allows an ontology of spacetime points.

% \subsection{Sophistication for diffeomorphisms: first pass}
In sum, sophisticated substantivalists characterize different possibilities as differing only qualitatively, and as being about spatiotemporal properties. Thereby, sophisticated substantivalists  endorse  \cite{EarmanNorton1987}'s condition of  \emph{Leibniz equivalence}: that isomorphic models represent the same physical possibility. And they moreover argue that  this understanding does not require a reductive, or eliminativist, picture:    isomorphic models are all equally valid representations of the same spatiotemporal  possibility.    In that sense, all spacetime distributions of the metric and matter fields that are related by a diffeomorphism are taken as representing the very same state of affairs (see also \cite[Sec. 3.3]{Pooley_draft} and both \cite{Pooley_routledge, Pooley_Read} for details).

%Here  I provide a brief summary and defense of sophisticated substantivalism, glossed in Section \ref{sec:intro} as a   position that maintains commitment to the existence of spacetime points and yet regards symmetry-related models as representing the same physical possibility. 

\subsection{Challenges to sophistication}\label{sec:soph_cheap2}\label{sec:skeptic}\label{subsec:soph_stab}\label{subsec:skeptic_why}
%Although something like sophistication has been suggested for gauge theories since its inception in terms of principal bundles (in \cite{YangMills}), only recently has the extension of the position from the context of general relativity  been more thoroughly conceptually analysed.  It was in  \cite{Dewar2017} that questions such as `When can sophistication be applied?', and `to what interpretative advantage?' started being seriously considered.  
 
 %To recap: in general relativity, the symmetries of the theory, given in equation \eqref{eq:equivalence}, are induced by the action of the diffeomorphisms on the spacetime manifold. In vacuum, these symmetries coincide with the isomorphisms of a well-understood mathematical structure:  (semi-)Riemannian geometry (cf. \citep{Oneill}). This coincidence,  perhaps aided by  our everyday acquaintance with space and time, helps us to accept the accompanying anti-haecceitist, qualitative or structural  interpretation of diffeomorphism-invariant quantities as being ``metaphysically perspicuous'' (as elaborated briefly in Section  \ref{subsec:soph_anti} and at greater length in Section \ref{sec:antihaec}). But many metaphysicians are not satisfied by the  sophisticated position, that claims anti-haecceitism is perfectly compatible with the existence of spacetime as a kind of substance.
 
 As they now stand, the considerations of Section \ref{subsec:soph_anti}, describing sophisticated substantivalism for spacetime, do not give us a general characterization,  for an arbitrary  theory, of what a sophisticated interpretation would be. In Section  \ref{sec:ext_soph}, I will, rather straightforwardly, extend sophistication to other theories. I will there follow \cite{Dewar2017}: a paper that kicked off a considerable recent literature. 
 
 % In Section  \ref{sec:soph_cheap} we will develop these ideas. There we will find, in the case of Yang-Mills theories, that  the same considerations apply, mutatis mutandis, with quidditism in place of haecceitism and properties in place of objects (or points).  The anti-haecceitist substantivalist  believes in the fundamental existence of space(time) points, but denies that these are metaphysically connected across worlds. 
But, even in the familiar case of diffeomorphism symmetry of general relativity,  the considerations of  Section \ref{sec:antihaec} should not be assumed to be unproblematic.  Indeed, many metaphysicians still find the anti-haecceitist interpretation of diffeomorphism symmetry opaque. And if we are to satisfactorily answer to these metaphysicians in the more general case, our answer should first be satisfactory for the more familiar case of diffeomorphisms. Thus, in order to set out the obstacles towards sophistication in the more general case,  we must first revisit the criticisms of sophisticated substantivalism. 

The main  idea of these criticisms is that stripping spacetime points  of primitive identity---condemning them to exist solely as clothed entities, dressed by the network of properties and relations in which they take part---is suspicious, or at least not capable of eliminating an ubiquitous and pernicious form of indeterminism (see \cite{Dasgupta_bare} and \cite{Teitel_hole} for recent examples).  Indeed, the label `sophisticated' was introduced with a deliberately negative connotation in \cite{belot_earman_1999} (see also \cite{belot_earman_2001}), signalling this  tension.\footnote{ The label has since been  accepted by many philosophers, who believe the tension is illusory. \cite{Dewar2017} extended the adjective ``sophisticated'' to other symmetries: where it is understood as allowing commitment to structure without reduction or elimination of the isomorphic copies. } 

Elaborating this  suspicion, \citet[p. 147]{Dasgupta_bare} issues two distinct challenges to the sophisticated substantivalist: ``(1) to clearly articulate what the underlying qualitative facts are like,
and (2) show that they are sufficient to explain (in the metaphysical sense)
individualistic facts about the manifold.'

 Without further context, these short quotes require some unpacking, and a good dose of interpretation. But in brief:    I see (1) and (2) as truly distinct;   I take (2) to be encompassed by what I labeled the  `worry about structure-tokens', whose treatment I leave  to the third paper, \citep{Samediff_1b}.\footnote{
 To see that (1) and (2)  are distinct,  suppose we intepret (1) such that a description of what the ``underlying qualitative facts are like'' is just the listing of the necessary and sufficient conditions for assessing qualitative identity of possibilities. Then, as I will argue in this Section, we can rest assured that sophistication, augmented by Desideratum (i) and (ii), answers this first challenge. But this is at best only a  criterion of (meta)physical identity of entire worlds; it may very well be insufficient  to `get inside' each world and characterize the individual facts qualitatively; thus (2) remains a separate challenge, and it is related to my worry about structure-tokens. 
  I will   return to  (2) %at the introduction to Section \ref{subsec:skeptic_why2} when I discuss the limitations of my response to  (1),  and
   more fully in the third paper in this series, \cite{Samediff_1b}.} And I take (1) to be, at bottom,    about the conceptual and physical obscurity of  the type of structure that remains invariant under the isomorphisms. Thus (1) is encompassed by what I labeled `the worry about structure-type', that I then fine-grained into Worries (1-i) and (1-ii) in the introduction to Section \ref{sec:soph_cheap}.

  And, in the same gist,  there are two recent debates within philosophy of physics, also corresponding to (1-i) and (1-ii). They are,  respectively: (1-i):  about whether  sophistication  is, in general,  suspicious, because it can be obtained all too easily, and thus (as  formulated  in the introduction to Section \ref{sec:soph_cheap}), it may leave the core invariant structure shared by the symmetry-related models as only implicit, or even obscure. And (1-ii): about whether the invariant structure---to which sophistication is metaphysically committed---is sufficiently `metaphysically perspicuous'.  Thus these suspicions, which I will describe more fully in Section \ref{par:cheap},  precisify   the metaphysician's Challenge (1). %At bottom, it is  about whether  the type of structure that remains invariant under the isomorphisms---the anti-haecceitist's qualitative ontology---is sufficiently clear.  

%In Section \ref{sec:responses} I will resolve these debates, and thus  answer the metaphysician's Challenge (1) as well. 

 %Jointly, these two replies will constitute an answer to the more recent debate, about whether sophistication is obtained all too easily. 

 %, and so it is not clear how t. %But in Section \ref{subsec:resp1} I argue that axiomatization rebukes this argument of the skeptic. 
%Then, {\old in Section}  \ref{subsec:skeptic_why}, the skeptic throws down a heavier gauntlet: that an axiomatization does not clarify \emph{individual} structural content. These challenges will be met in Section \ref{sec:gr_re}. 

\subsubsection{Extending Sophistication and invariant structure}\label{sec:ext_soph}

 To clarify the extension of sophistication to a wider class of theories beyond general relativity, \citet[p. 502]{Dewar2017} contrasts it with \emph{reduction}, i.e. what I called  `eliminativism' in Section \ref{sec:intro}:  eliminativism (or reduction)  advocates
a new theory that excises the structure that could distinguish the  isomorphic models, so that symmetry transformations would fix, i.e. act as the identity on, models of the new, reduced theory. On the other hand, models of a sophisticated theory are acted on by, but are to be isomorphic under, symmetry transformations. And that is it: that is the extension. 

When we leave the concrete examples of general relativity and Yang-Mills theory, this characterization of sophistication seems rather unsubstantial.  Thus the first accusation that `sophistication is gotten on the cheap' is that declaring, by fiat, that symmetry transformations are isomorphisms is too easy;  Russell's `theft over honest toil' comes to mind. 

For, according to Dewar, the requirement that symmetry-related models be isomorphic may not be stringent at all:  whenever a theory has symmetries, we could extend  sophistication, even to cases where those symmetries do not correspond to any of the better known examples of mathematical structure, by altering the semantics   of that theory, such that
``we obtain [the new semantics] by taking the old objects,
and declaring, by fiat, that the symmetry transformations are now going to ``count''
as isomorphisms''. Thus Dewar  proposes that given  a space of models and a symmetry transformation acting on this space, one could  just announce an invariant  structure as defined implicitly by `whatever is left invariant' by the action of the symmetries.  In such cases, knowing the isomorphisms sheds no light on what structure they are leaving invariant. 

Dewar calls this extension the  `external method' \cite[p. 502]{Dewar2017}. Here is \citet[p. 502]{Dewar2017}:\begin{quote}
Is there a way to precisify what is meant?
Here is one way to do so. Rather than trying to define the objects of the new
semantics ‘internally’, as mathematical structures of such-and-such a kind (paradigmatically,
as sets equipped with certain relations or operations), we instead
define them ‘externally’: as mathematical structures of a given kind, but with
certain operations stipulated to be homomorphisms (even if they’re not ‘really’
homomorphisms of the given kind). For example, one way to define vector spaces
is to define them as sets equipped with operations of addition and scalar multiplication,
obeying appropriate axioms. This is the internal method. The alternative
is to define them as spaces of the form $\RR^k$, with the further feature that linear
transformations are declared to be homomorphisms—and in particular, that invertible
linear transformations are isomorphisms. This is the external method.\end{quote} 
To further explicate the external method, and contrast it with the internal,  it is useful to follow \cite{Klein_erlangen}'s  distinction, taken as a starting point for the Erlangen programme. That distinction  is reflected in \cite[p. 502]{Dewar2017}'s jargon  of  `external vs internal', but it is made more precise in \citet[Ch.  4.1]{Jacobs_thesis}, as:\\
\noindent \textit{The symmetry-first approach (external)}: finding a structure  implicitly  as `what stays constant across the symmetry-related models';\\
\noindent \textit{The structure-first approach (internal)}: finding the symmetry-related models as those that possess the same structure.

Thus far in our treatment of symmetries (see \cite{Samediff_0}), we have been given some space of models of a theory and a notion of dynamical symmetry: so to use Jacobs' jargon, we were adopting the symmetry-first approach---so that, at least at first sight, we ``could only glimpse'' the structure thereby defined as whatever is invariant under the symmetries. In the structure-first (or internal) approach, the aim is to give a characterisation of
a structure in terms of relations defined over its domain, such that this structure
is invariant under the dynamical symmetries of the theory.  As Jacobs puts it: \begin{quote}
Structure-first sophistication consists of: the
stipulation of a set of relations over the theory's sub-domains, such that the bijections
which induce dynamical symmetries of the theory's models leave these relations
invariant. 
\cite[p. 62]{Jacobs_thesis} \end{quote}

\subsubsection{Sophistication on the cheap?}\label{par:cheap}

 We are here in the vicinity of two philosophical debates about symmetry, that, to my view, add further detail to the metaphysician's challenge  (1), broadly about `the obscurity of structure-type', that we glimpsed in the case of general relativity in the introduction to Section \ref{subsec:soph_stab}.% (broadly stated:  to clearly articulate what the underlying qualitative facts are like.) 

The first debate was sketched in Section \ref{sec:ext_soph}, and is about whether sophistication can be attained too easily, i.e. whether it thus needs further explication. For sophistication by brute force, i.e. Dewar's `external' sophistication (and also Jacob's symmetry-first approach), seems liable to leave, even at a strictly mathematical level, invariant structure as opaque, or not sufficiently understood. In the introduction to Section \ref{sec:soph_cheap}, I related this debate to Worry (1-i).

And this lack of clarity about the common mathematical structure motivates the debate about whether symmetry-related models of a given theory should invariably be regarded as being physically equivalent, i.e. as making the same claims about all physical facts about the world, even in the absence of  a clear, or  `metaphysically perspicuous', understanding of the common ontology of the models. In the introduction to Section \ref{sec:soph_cheap}, I related this debate to Worry (1-ii).

\cite{Moller} introduced a helpful labelling for the two sides of this second debate on physical significance. He labels  the more permissive answer---symmetry-related models are invariably physically equivalent---as the \emph{interpretational approach} to symmetries;  and he contrasts it with  the \emph{motivational approach}, which requires a characterisation of the
common ontology of symmetry-related models before acknowledging physical equivalence. This is the answer he endorses.

 %Thus demanding reduction is too strict, and  we must keep searching for an answer to the second debate---about when sophistication sufficiently clarifies underlying structure---in order to decide our answer to the first debate: about (what amounts to) a metaphysically perspicuous package for the symmetry-related models.  

As to the first debate,  Section \ref{sec:ext_soph} reported on an initial attempt to repel the accusation of cheapness, by distinguishing internal from external sophistication. But as we will now see, that attempt is not satisfactory. 

All hands agree (e.g.  \citet{Moller, ReadMartens, Jacobs_thesis}) that an implicit definition, as in Dewar's external method, would not sufficiently explicate the common ontology of the symmetry-related models. According to \citet[p. 1264]{Moller}: ``[I]t is simply opaque what, according to the external approach,
the world is really like.'' 
Thus
\citet[p. 9]{ReadMartens} endorse \citet[p. 502]{Dewar2017}'s `internal'  sophistication:  equivalence can only be justified in cases where the symmetries  coincide with isomorphisms \emph{of some familiar structure}.

But the challenge here is to characterize just what `familiar strucures' are. For the role of familiarity is to stop the externally defined symmetries from being reconstrued as internal by just declaring the invariant structure to be whatever is preserved by the symmetries. This may seem tautologous, or at least a mere reconstrual, but it nonetheless threatens to collapse the distinction between internal and external. Thus far, being internal rather than external seems to rely on whether the structure is `familiar' or not: a vague and conservative criterion that does not  illuminate the question. % Here our desideratum for a conceptually `clear' definition of structure may thus outstrip the purely mathematical discussion, elaborated on Section \ref{sec:structuralism_maths}. And though  the desideratum is no longer strict, it seems either trivial or at least too opaque to satisfy the skeptic about structuralism. 
%\footnote{ This broad idea works out beautifully for the example that Jacobs focuses on (scaling of masses in Newtonian gravity). In that case, one successfully characterizes the mathematical structure of the theory first, and then deduces the isomorphisms that preserve that structure. At least in that example, it seems that sophistication is apparently \emph{not} condemned as cheap.}

It is helpful to discuss these issues in the supposedly clear case  of general relativity. Even there, we have not yet  precisely identified \emph{the type of structure} that is invariant. The idea  is that the models of the theory are structured sets, $\mathcal{D}=(D, R)$, consisting of a base (unstructured) set $D$ and relations among the elements of this set, which we can here give just one label, $R$, which are supposed to represent the invariant structure.  
 In general relativity, a model ``consists of a base manifold $M$ over which
we have defined some (geometrical) structure in the form of the tensor fields'' \cite[p. 60]{Jacobs_thesis}.  But what exactly are the relations that stay invariant under the symmetries, i.e. which stay invariant under the (pullback of) active diffeomorphisms? Tensor fields certainly do not remain invariant.  A statement such as ``the metric tensor is $g_{ab}$'' is not symmetry-invariant; it can be true in one  model and false in a  symmetry-related one. The distances between the points of $M$ seen as an unstructured base set are also not invariant, since the distance between $x$ and $y$ according to $g_{ab}$ is not the same distance as that  according to the dragged-along metric $f^*g_{ab}$. The same reasoning would of course apply to angles, Riemann curvature scalars, etc.\footnote{And in particular, I don't find the discussions in the standard textbooks of general relativity illuminating on this issue. Here is \citet[p. 438]{Wald_book}:``If a theory describes Nature in terms of a spacetime manifold $M$ and tensor fields, $T$, then if $f:M\rightarrow N$ is a diffeomorphism, the solutions $(M, T)$ and $(N, f^*T)$ have physically identical properties. Any physically meaningful statement about $(M, T)$ will hold with equal validity for $(N, f^*T)$. On the other hand, if $(M, T)$ and $(N, T')$ are not related by a diffeomorphism, \emph{and if the tensor fields $T$ represent measureable quantities}, then $(N, T')$ will be physically distinguishable from $(M, T)$." (my emphasis). }

 The conclusion is that even if we try to apply the general, straightforward  definition of internal sophistication---as a property of structured sets---to the reasonably well-understood case of general relativity, the structure that remains invariant under isomorphisms is characterized as the abstract set of  quantities that are invariant under pull-back: just as the symmetry-first, or external, approach---not the structure-first, or internal, approach---would suggest.\footnote{Besides the equivalence class of metrics,  the smooth structure of the manifold stays invariant under arbitrary diffeormorphisms. That is because it is defined by a maximal atlas, which is already an abstract equivalence class, for all of the diffeomorphism-related  charts (see \cite[Sec. 5]{Samediff_0}).}  
 
 In Section \ref{sec:responses} I will overcome these shortcomings about the usual definition of internal sophistication, and thus resolve Worry (1-i).  My Desiderata (i) and (ii) are the core of my response, since they give further criteria for the invariant structure, that go beyond its mere familiarity.  %\footnote{I should note that Jacobs does not condemn the symmetry-first approach as  `a metaphysical black box', since it provides  a decision procedure for determining whether a quantity or relation should be considered real (namely, invariance). Nonetheless, he acknowledges that there is an important difference between this and providing a fundamental set of invariant relations, as in the internal approach.}  %Thus the qualification, in Section  \ref{par:active_passive}, that ``there is no known, \emph{non-trivial} way to get rid of diffeomorphism symmetry''.

Moving on to the second debate, corresponding to Worry (1-ii): even supposing we have succeeded in defining some structure internally in a non-tautologous manner, there is still the matter of connecting the mathematical and the physical. Given the vast and flexible resources of mathematics, a strictly formal internal definition will hardly quell the skeptic's suspicion that the said structure fails to support the metaphysical interpretation we would like to hang on it. As argued by \cite{ReadMartens}: if there is no `metaphysically perspicuous' package accompanying the formulation of the symmetry-related models, sophistication is still metaphysically obscure.  

Thus, with this second debate in mind, following a nomenclature suggested in \cite[p. 220]{Belot2003},  \cite{Moller}  (see also \cite{ReadMartens}) insists we still need to find a  metaphysically perspicuous interpretation of the invariant structure, even in the internal case.\footnote{\cite[p. 220]{Belot2003} talks about finding a `perspicuous formulation' of the symmetry-related models, \cite{Moller} also talks about perspicuous characterization.}  Thus,  \citep[p. 9]{ReadMartens} advise, even for internal sophistication, that more needs to be said so that we can  ``regard those models---interpreted as being isomorphic---as in fact representing the same physical states of affairs.'' %And while  we cannot,  using only local fields or with reference to the underlying points of $M$, explicitly denote this structure without redundancy,  the isomorphisms and symmetries do \emph{not} need to be eliminated from the theory  in order for the structural interpretation to succeed; p
And thus there is a contradition  with \cite[p. 62]{Jacobs_thesis}, who writes: 
\begin{quote}
 If we agree that an interpretation of a theory provides a metaphysically
perspicuous picture if it tells us plainly and clearly which entities the theory posits
(ontology) and what the fundamental relations between these entities are (ideology),
then structure-first sophistication is perspicuous in this sense.
\end{quote}
 %{\old; in Section \ref{sec:soph_cheap}, we will try to provide such desiderata.}
  I agree with \cite{Moller, ReadMartens}: even assuming we have obtained a non-tautologous internal  characterization of structure,  we won't necessarily have succeeded in providing a metaphysically perspicuous picture, because even the internal picture may not yet state ``plainly and clearly which entities the theory posits
(ontology)''.

But at this point,  these requests for `metaphysical perspicuity' remain vague: so far,  `perspicuity' seems to be  perhaps best understood  as `intelligibility' of the mathematical structure. As with the label of `internal', the label of `metaphysically perspicuous' does not yet satisfy any rigorous requirements.

 In Section \ref{sec:responses} I will also respond to this second debate/Worry (1-ii).  My novel proposal is to characterize perspicuity as part of my Desideratum (ii), by explicitly anchoring the mathematical structure of the theory on axioms bearing primitive physical significance. 
 
After this review of the debates/worries about the `obscurity of structure-type', i.e. (1-i) and (1-ii),  that precisify Dasgupta's Challenge (1) to  sophistication, I now turn to my response.

\section{Sophistication defended}\label{sec:responses}

In this  Section,  I will provide some answers to the debates reviewed in the preceding Section and to the metaphysician who is skeptical about sophistication. They are, now in a bit more detail than provided in their formulation in the introduction to Section \ref{sec:soph_cheap}: \\
\indent (1-i) Whether the invariant structure is sufficiently (internally) understood, and whether it refers to some notion of `qualitative' property;\\
\indent (1-ii) whether the invariant structure is the only part of the theory that refers, i.e. whether it is `metaphysically perspicuous'. \\
And the metaphysician's concerns are just a coarse-graining of these worries. To recap: he  may  not be satisfied, for instance, that the common ontology of two isomorphic models of general relativity  is just `fields on the manifold, where the latter is interpreted anti-haecceitisically'. He  is concerned that a characterization of structure as `qualitative' comes all too cheaply, as a bare intuition that cannot be (meta)physically cashed out. %According to him, this understanding should be supported   by an independent and metaphysically perspicuous characterization of structure, which is not necessarily supplied by sophistication.
 % Plausible resolutions of this concern should respond to the threat of  indeterminism; the main  physical motivation for an analysis of the ontological status of symmetry-related models.
%And the common idea of \cite{Dewar2017, ReadMartens, Jacobs_thesis} is that we can escape the accusation of cheapness in two steps: (i) insisting symmetries  coincide with isomorphisms of some structure that is given `internally', and (ii) first defining the structure, and then finding the symmetries/isomorphisms that preserve it. We will satisfy these demands in Section \ref{subsec:resp1}.% (1) is a methodological principle, but (ii) requires some groundwork, and, as we will see, even once achieved it may not fully satisfy a skeptic about structuralism.

%We begin in Section \ref{subsec:resp1} by furnishing the mathematical desiderata for sophistication of symmetries. In Section \ref{sec:} we provide the physico-mathematical desiderata. 
 
% \subsection{Mathematical desiderata for sophistication}
Before   I propose my  two desiderata for sophistication (now in more detail), we need some definitions (for more information about the nomenclature here, see \cite[Sec. 2]{Samediff_0}). First, let $\F$ be the space of models of the theory  whose models $\varphi\in \F$ are maps from a tensor bundle of valence $(r,s)$ over a structured, finite-dimensional  base set $N$, to a value space $F$, i.e.: $\varphi: T^{r,s}N\rightarrow F$. Now we can define three different notions of symmetry acting on $\F$.

Let  $\G$ be the group of isomorphisms of $\F$, so that $\F$ is a groupoid, i.e. a category in which every arrow is an  isomorphism, with the objects of the category being the models, i.e. the elements of   ${\F}$ (cf. \cite[Sec. 2.3]{Samediff_0} and footnote 15 therein for more on this):   
\begin{eqnarray}
\G\times \F&\rightarrow& \F\nonumber\\
(g, \varphi)&\mapsto& \varphi^g\,\, .\label{eq:G}
\end{eqnarray}

Let  $\mathsf{Aut}(N)$ be the group of automorphisms of $N$, that preserve whatever structure of $N$ that we are holding fixed (e.g. smooth, fibered, product, etc). Mathematically,  any given notion of automorphisms of the base set  can be pulled-back to define a corresponding transformation on the (values of the) fields, which can be an isomorphism in the relevant category that has the theory's models as objects.\footnote{As described in \cite[Sec. 4]{Samediff_0}: the Ehresmann connection $\omega$ is as described above, a map from the tangent bundle of the principal bundle to a Lie algebra, i.e. $\omega:TP\rightarrow \mathfrak{g}$, and the metric is a map from the symmetric tensor product of the tangent bundle to the reals $g_{ab}:TM\otimes_S TM\rightarrow \RR$. What matters for this discussion is that an action of the automorphisms of $P$ or $M$ can be `pulled-back' to an action on the values of these fields.  }  Therefore $\mathsf{Aut}(N)$ acts on $ T^{r,s}N$ and thus acts through pull-back on $\F$, as  (cf. \cite[Sec. 3]{Samediff_0})
\be \mathsf{Aut}(N)^*:\F\rightarrow \F \label{eq:Aut}\ee
Now, the automorphisms of $N$ cannot change the relation between the fields on $N$ and $N$'s  fixed background structure, since this structure is  invariant under the action of $\mathsf{Aut}(N)$. Therefore we can interpret $\mathsf{Aut}(N)$ to only change  `which object plays which role'.  And so we can regiment the idea of qualitative with the following definition:
\begin{defi}[Qualitative]\label{def:qualitative} Given, $\F$ and $N$ as defined above, those properties  that remain invariant under $\mathsf{Aut}(N)^*$, that is, invariant under the induced action of the automorphisms of the (structured) base set, will be called \emph{qualitative} properties of $\F$.\footnote{Note that this definition is not quite the same as the `verbal characterization', given in the second paragraph of Section \ref{subsec:soph_anti}, that referred to proper names, indexicals, etc. If we take coordinate systems to provide names for the objects of our manifolds (which is not the usual understanding of `names' in this context) then Definition  \ref{def:qualitative} would be analogous to the verbal characterization \emph{provided} we impose (standard) mathematical restrictions  on our `naming'  the objects of a manifold \emph{and} we have an active-passive correspondence for the automorphisms of $N$. This was shown to hold for smooth manifolds and fibered manifolds in \cite[Sec. 5]{Samediff_0}.\label{ftnt:qualitative}}\end{defi}

And finally, let $T$ be a dynamical theory for $\varphi\in \F$, with $\bar \G$ as its group of dynamical symmetries, i.e.:
\begin{eqnarray}
\bar\G\times \F&\rightarrow& \F\nonumber\\
(\bar g, \varphi)&\mapsto& \varphi^{\bar g}\,\,.\label{eq:barG}
\end{eqnarray}
 These symmetries are here restricted to preserve the value of a salient set of quantities (usually the action functional or the Hamiltonian), and to preserve some background structure of $\F$ (for example, the symplectic structure; see \cite[Sec. 2.1]{Samediff_0} and Definitions 1 and 2 therein for more on this).

Now note that according to Definition \ref{def:qualitative}, 
qualitative properties are independent of the dynamics of the theory $T$. Our first Desideratum makes sure that all three notions of `symmetry', given in equations  \eqref{eq:G}, \eqref{eq:Aut}, and \eqref{eq:barG}, match.  And out of the three notions, the second desideratum concerns only   the isomorphisms,  \eqref{eq:G}: it is a condition on how this group should be constructed. 
 
 The two Desiderata are:
 \begin{enumerate}[(i)]
  \item Suppose theory $T$ is given, jointly with $\F,\G$ and  $\bar \G$. The isomorphisms $\G$ are required to be:\\
    (i$_a$) such that isomorphisms and  symmetries coincide, $\bar \G=\G$ (i.e. the action of  $g\in \G$ and of $\bar g\in \bar \G$ on each element of $\F$ correspond one-to-one)
   and, moreover,\\
    (i$_b$) induced by the   automorphisms of some base space: 
  \begin{align} \text{For all}\,\,\varphi\in \F,\,\,\text{and for each}\,\, &g\in \G,\,\, \text{there exists a unique}\, t_g\in \mathsf{Aut}(N)\,\,\text{such that}\nonumber\\
 ~& \varphi^g=t_g^*\varphi\,\,;
  \end{align}
  \item   $\F$ and $\G$ are not initially given, but jointly arise from: (ii$_a$)  an axiomatization; that (ii$_b$) uses only terms that  have a direct physical interpretation.  
 \end{enumerate}
% Thus, item (i) guarantees that anti-haecceitism/anti-quidditism suffices to respond to the threat of physical indeterminism  in each theory; and item (ii) guarantees that the invariant structure has a metaphysically perspicuous interpretation as an internally sophisticated theory. In more detail,

In  Section \ref{sec:cr1} I will show that my Desideratum (i) gives at least a partial answer to Worry (1-i) by conjoining the arguments of this paper with some of the results of \citep{Samediff_0}. Desideratum (i) first of all guarantees a necessary, but not sufficient condition for the interpretation of isomorphic models as physically identical: namely, that their differences are physically unobservable. The argument here, articulated in \cite[Sec. 2.3]{Samediff_0} (based on an argument of \cite{Wallace2019a}),  is that if observations are dynamical processes, then  dynamical symmetries leave all observations invariant, and thus are unobservable.  By requiring the isomorphisms of the models of the theory to be identical to the dynamical symmetries of the theory, Desideratum (i) guarantees that isomorphisms are physically unobservable.

 Second,  Desideratum (i) requires the isomorphisms to be induced by the automorphisms of the underlying structured base set of the theory, which I will take to mean that   these automorphisms affect only `which objects play which role', in line with Definition \ref{def:qualitative}. 
 Thus Desideratum (i)  guarantees that the differences between isomorphic models are \emph{not} qualitative, in a well-defined notion of qualitative. 
 
Note that the actual relations and quantities that remain invariant under these automorphisms are still only implicitly defined, i.e. by the symmetry-first method. Nonetheless, by  constraining  the notion of isomorphisms, Desideratum (i) protects the definition of internal sophistication from `cheapness', i.e.  from being mere (internal) reconstruals of arbitrary (external) symmetries.\footnote{The third Desideratum, to be described in \cite{Samediff_1b},  requires those automorphisms to have a passive correspondence as a change of physical coordinates. Then,   if both the Desideratum is satisfied and an active-passive correspondence for the symmetries exists,  like the one shown to exist for diffeomorphisms and gauge transformations in \cite[Sec. 5]{Samediff_0}, we can  construe the differences between isomorphic models as merely notational. Jointly, these arguments give an interpretation of qualitative as `coordinate-independent' as opposed to the vaguer, but similar in spirit, `without the use  of singular terms'. \label{ftnt:unnoficial}}

 This remaining concern about a `structure-first definition' will only be resolved in Section \ref{sec:cr2}. There I will show that my Desideratum (ii) answers challenge (1-ii), by requiring the invariant structure to have a primitive physical significance, since it is to be defined axiomatically  in terms of primitive physical terms. I will verify that this resolution applies to spacetime metrics and Ehresmann connections in principal bundles. The idea then is that other theories will be sophisticated in the appropriate way if they also admit such constructions.  
  
   %The perspicuity of the interpretation hangs on whether the axioms used can be understood as basic physical posits of the theory at hand. 

 %This Section can therefore be taken to answer the debate about sophistication alongside one of the questions from the metaphysician, represented by \cite{Dasgupta_bare}'s   Challenge (1): ``to clearly articulate what the underlying qualitative facts are like'' (described in Section \ref{subsec:skeptic_why}).\\

 % But it is important to highlight the limitations of this resolution: I do not take it to answer another sort of  question, represented  by  \cite{Dasgupta_bare}'s  Challenge (2): ``to show that the underlying qualitative facts are sufficient to explain individualistic facts about the manifold''.  But I will not address that Challenge in this paper; this will be left for  the third paper in the series, \cite{Samediff_1b}. Here I will merely further expound it, in Section \ref{subsec:skeptic_why2}. 
%In the second paper in the series, \cite{Samediff_1b}, I will respond to Challenge (2), by using particular relational bases for piecewise qualitative facts, thereby showing that spacetime counterfactuals can be articulated via a theory of  spacetime counterparts.
 
 %To answer this second challenge, we need to `get inside' the models, and  provide qualitative descriptions of points or regions. This of course goes beyond the specification of the type of structure at hand: it requires particular descriptions of structure tokens as well.  We will return to this problem in Section \ref{subsec:skeptic_why2}.

 \subsection{Desideratum (i): symmetries induced by automorphisms of a base set}\label{sec:cr1}\label{subsec:resp1}
 My first desideratum  is essentially the content of Earman's two `SP principles' about spacetime symmetries, proposed in \cite[p. 45-47]{Earman_world} and often taken to be extendible to gauge theory (see e.g. \cite[p. 9]{Belot_sym}, \cite[Ch. 4, 8]{Jacobs_thesis}, \cite{Jacobs_PFB, Hetzroni_gauge}). Jointly, the two principles require that the dynamical symmetries should  be just those induced by automorphisms of the structured base set (cf. \cite[Footnote 14]{Samediff_0} for my definition of automorphism and the previous Section, for my assumptions of a base set for the dynamical variables). In the case of general relativity, this structured base set is a smooth manifold, and, in the case of Yang-Mills theory, it is a fibered smooth manifold called a principal bundle. 
 
  To articulate the principles, Earman distinguishes  internal and external parameters. External parameters are the independent, or base-set variables, i.e. the points of $N$ in the previous Section, and thus correspond, in our two main study-cases,  to spacetime points or to the points of the principal fiber bundle (as organized in orbits); internal parameters are  the value spaces $F$---where the  fields  of different (field) theories take their values, taking as their argument  the independent, external or base-set variables.    %And Earman's  SP principles jointly require that the dynamical symmetries of spacetime theories  coincide with this induced action of the automorphisms of the base set. 
  
  In more detail, Earman's SP-principles are, in a language suitable for this paper:\smallskip\\\smallskip
\noindent \emph{SP1}: Any  symmetry  of theory $T$ is induced by some automorphism of spacetime $\mathsf{Aut}(M)$, where $M$ is spacetime, i.e. 
\be \text {Given}\,\, \bar g\in \bar \G, \quad \varphi^{\bar g}=t^*\varphi\quad \text{for some}\,\, t\in \mathsf{Aut}(M);\ee\\
\noindent \emph{SP2}: Any  automorphism of spacetime yields a   symmetry of theory $T$, 
\be \text {Given}\,\,  t\in \mathsf{Aut}(M), \quad t^*\varphi=\varphi^{\bar g}\quad \text{for some}\,\,\bar g\in \bar \G.\ee

I take this to be the content of the  principles (in the original context of spacetime): they  ensure a theory's  spacetime has  `just the right amount' of structure for symmetry-related possibilities to be identified by an anti-haecceitist. 

That is,  if \emph{SP1} fails then there is a symmetry of $T$ that does not correspond to an automorphism of spacetime.  For example, following the example of Section \ref{subsec:soph_anti}, of Newtonian mechanics, we saw that spacetime contains
a standard of absolute rest: spacetime is a geometric structure, $\bb E^3\times \RR$ (cf. \cite[p. 9-12]{Earman_world}), whose automorphisms do not include Galilean boosts. Thus for $\bar g$ being a boost, we can find no corresponding $t\in \mathsf{Aut}(\bb E^3\times \RR)$. Accordingly, `being at rest' is a qualitative fact in Newtonian mechanics.  On the other hand, Newtonian mechanics does not have a standard of absolute {\em place}, e.g. preferred origin. Thus for $\bar g$ being a uniform translation of particle positions, there will always be a $t\in \mathsf{Aut}(\bb E^3\times \RR)$.

According to Definition \ref{def:qualitative}, anti-haecceitism---and more specifically, the sophistication I am defending---will identify   possibilities related by static shifts, i.e. spatial and temporal translations, but not by kinematic shifts, i.e. boosts; see \cite{MaudlinBuckets} and \cite{sep-newton-stm} for more about this distinction (see also footnote \ref{ftnt:abs_vel} in Section \ref{subsec:soph_anti}).
That is,  boosts are dynamical symmetries of Newton's laws (which comprise the theory $T$), which
means that the standard of absolute rest is irrelevant to the theory's dynamics---and SP1 fails. In short,  anti-haecceitism, or the denial of non-qualitative facts, does not account for this dynamical symmetry of T.\footnote{Note, however, that if we formulate the dynamics of non-relativistic particles through an action principle, then boosts will not preserve the value of the action: they add boundary contributions. Thus, in Newtonian mechanics, boosts are not dynamical symmetries according to my primary desideratum here. } 

Accepting \cite{Moller}'s Motivationalism, we should feel `motivated' to find a new theory (e.g. Neo-Newtonian, or Newton-Cartan gravity), for which the dynamical symmetry \emph{is} induced by an automorphism of a natural geometric structure. 

Conversely,
if   \emph{SP2} fails, then there is at least one automorphism $t\in \mathsf{Aut}(M)$ of spacetime that is not a dynamical symmetry $\bar g\in \bar \G$ of $T$. 
So for our main case of spacetime and general relativity, and haecceitism vs anti-haecceitism about points: a failure of SP2 would involve the theory T, i.e. general relativity, assigning a dynamical role to spacetime points and regions. That is, two models, differing only by which region supports which pattern, would not be related by a dynamical symmetry, and thus  would be empirically discernible (according to the converse of the  empirical unobservability thesis, discussed in \cite[Sec. 2]{Samediff_0}). 
So theory $T$ would provide a dynamical role for singular terms of spacetime, and thereby draw empirical distinctions between regions of spacetime which are qualitatively identical. 
Thus, in this case, an anti-haecceitist position about spacetime would have to rub out empirical  differences.

These ideas and specific principles SP1 and SP2 are straightforwardly extendible to gauge theories (see e.g. \cite[Ch. 4, 8]{Jacobs_thesis}, \cite{Jacobs_PFB, Hetzroni_gauge}). % Again assuming symmetry-related models are the only physically identical models (see \cite[Sec. 2]{Samediff_0}), Desideratum (i)  guarantees that, given the space of models, only those related by an isomorphism can represent identical physical possibilities. Moreover, this desideratum also guarantees that these isomorphisms, by being induced by the automorphisms of a base space, reflect only changes of which points or regions support which patterns and relations. Thus there is a sense of `qualitative' that is well translated by anti-haecceitism/anti-quidditism, and that characterizes each physical possibility. 
%Given a space of models $\F$ of some dynamical theory T, that describes values on some base set $M$, i.e. $\varphi\in \F$ is a section of some vector bundle over $M$ (cf. \cite[Sec. 3]{Samediff_0}). 
Now we can make the argument  sketched in the introduction of this Section \ref{sec:responses} (penultimate paragraph), more like a deductive argument:\\
 \indent Assumption 0: The physical differences between worlds modeled by $\varphi_1, \varphi_2$ are `in principle' observable.\footnote{This assumption is far from innocuous: it identifies physical differences,  empirical differences, and observable differences.  But here I do \emph{not} use `empirical' to denote the traditional positivist and post-positivist `meter-readings' or `no-special-training-neeed for the judgment', or `the sheer look'---a very common denotation in the  literature about the theory-observation distinction of the past fifty years. I use it to denote `in-principle-observable', in a very encompassing sense of `in-principle', that relies only on the fact that observations are dynamical processes (see  \cite[Sec. 2.2]{Samediff_0}). Of course, a haecceitist might very well, and consistently, believe that haecceitistic differences exist but are not empirical. In the present nomenclature they would be only `metaphysical'.}  \\
  \indent Assumption I (Unobservability): Two models $\varphi_1, \varphi_2$ are observationally equivalent iff they are related by a dynamical symmetry $\bar g\in \bar \G$ of $T$ of \eqref{eq:barG}, i.e. iff $\varphi_2=\varphi_1^{\bar g}$;\footnote{Note that this symmetry need not respect the fibers of the vector bundle: e.g. though the metric is a section of a vector bundle over spacetime, diffeomorphisms do not preserve the fibers. }\\
\indent Assumption II (Desideratum ($i_a$)): Dynamical symmetries coincide with  the   isomorphisms of the space of  models of the theory, $\G=\bar G$;\\
\indent Assumption III (Desideratum ($i_b$)): the isomorphisms $\G$ of the space of models $\F$ are induced by the automorphisms of the base set of $\varphi\in \F$, i.e.  $\mathsf{Aut}(N)$.\\
 % Assumption III:  Physical possibilities are individuated qualitatively;\\
 \indent Conclusion: for $T$ obeying Desideratum (i),  the threat of physical indeterminism  is repelled by a commitment to a qualitative ontology. \\
  \indent Proof: the threat of physical indeterminism  arises from the existence of dynamical symmetries. The threat is dissolved if we are to physically identify symmetry-related possibilities. This is how far Assumptions 0 and I get us, as argued in  the last paragraph of \cite[Sec. 2.2]{Samediff_0}, on \emph{empirical unobservability} of dynamical symmetries.  Assumption II translates dynamical  symmetries to a notion of mathematical isomorphism, as described in \cite[Sec. 2.2]{Samediff_0}. The automorphisms of the base set $\mathsf{Aut}(N)$ are those permutations   that preserve some base set structure (smooth, fibered, etc): under these constraints,  they change the `which point is which', or `which property is which' (according to Definition \ref{def:qualitative}). And so we take these automorphisms to give only non-qualitative differences; thus by Assumption III,  symmetries give only non-qualitative differences, and we obtain the conclusion.
  
%In other words,  this desideratum guarantees that the isomorphisms of the theory reflect only changes of which points or regions support which patterns and relations. Thus there is a sense of `qualitative' that is well translated by anti-haecceitism/anti-quidditism, and that characterizes each physical possibility. 

%; knowing that qualitative facts are those invariant under isomorphisms will again not help in this endeavour.}

 %This is the challenge of finding a perspicuous metaphysical intepretation of an internally sophisticated theory: one in which the structure is not deduced by first postulating the space of models and some notion of isomorphism therein. Desideratum (ii), to be developed  in  Section \ref{sec:cr2} below,  places the space of models and their isomorphisms downstream from an axiomatic, structure-first definition, that uses basic premises with physical content. %and indeed \cite{Jacobs_thesis}'s imposes this extension as the main desideratum for internal sophistication.
 %Plausibly, a complete defense of anti-haecceitism and anti-quidditism involves a defense of Earman's principle.In the follow-up \cite{Samediff_1b}, I will argue that a further justification for this desideratum is that it allows a deflation of symmetries by giving them a passive gloss. That is, once symmetries are construed as isomorphisms that are  induced by the  automorphisms of a natural geometric structure, we can build an active-passive correspondence for the isomorphisms that allows us to understand symmetry-invariance as mere invariance under coordinate changes. 

We have already shown that the  symmetries of  general relativity and  Yang-Mills (in the principal bundle or in the Atiyah-Lie bundle formalisms) can be understood as isomorphisms induced by the automorphisms of an underlying geometric structure, in \cite[Sec. 5]{Samediff_0}.  So both theories fulfill Desideratum (i).  What is left to say? 

Assumptions II and III above take the isomorphisms of the models, the dynamics, and the automorphisms of the base space as given. Thus Desideratum (i) regiments the notion of `internal sophistication' by relating these three  notions, thereby severely constraining the structure that is to be left invariant. But the skeptic could still be unsatisfied: even though Desideratum (i)  avoids cheap sophistication, it does not provide a truly  `structure-first' definition of the invariant structure.  Desideratum (ii), to be treated next in Section \ref{sec:cr2},  will fill this gap: isomorphisms will emerge from an axiomatic, or rather, synthetic, definition of the invariant structure; that is, the isomorphisms will emerge from giving the structure first.\\

 Before we leave this Section, I want to connect it with both what has come before, in \cite{Samediff_0}, and what  is to come, in \cite{Samediff_1b}. As argued in  in \cite[Sec. 5]{Samediff_0}, fulfillment of Desideratum (i)  in the case of Yang-Mills theory and general relativity also allows a deflation of symmetries by giving them a passive gloss (see also footnote \ref{ftnt:unnoficial}). That is, once symmetries are construed as isomorphisms that are  induced by the  automorphisms of a structured base set, and we can build an active-passive correspondence for those  automorphisms, we can   understand symmetry-invariance as mere notational invariance. Thus, for theories that satisfy Desideratum (i), we are motivated to construct a passive-active correspondence, which would indeed reinforce the common belief that the symmetries of those theories correspond to notational redundancy. Such a construction is the basis of our Desideratum (iii), to be expounded in \cite{Samediff_1b}. 

 \subsection{Desideratum (ii): axiomatization of the invariant structure}\label{sec:cr2}

In this Section, I will put the final nail on the coffin of Worry (1) (and alongside it, of the metaphysician's Challenge (1)) of Section \ref{sec:soph_cheap2},  at least for the well-understood cases. 

 In Section \ref{subsec:skeptic_why} I have already argued that previous descriptions of sophistication may lack that which Desideratum (ii) would guarantee: a metaphysically perspicuous interpretation as an internally sophisticated theory. This Section will answer  these two related concerns, viz. about a perspicuous metaphysics and  about the isomorphisms being of some invariant structure that is understood first (or internally, but in a more robust sense than that provided by Desideratum (i)).%---there, the invariant structure was  understood via the automorphisms of a structured base space. 
%Moreover, the metaphysician's concern above also requires a precisification of `perspicuity'. 
%In these terms, we have cast  doubt on the supposedly well-understood case of general relativity; and so  we cast doubt also on the anti-quiddist interpretation of gauge theory, and of more general sophisticated theories. 
%, by characterizing the underlying qualitative facts axiomatically. 

Here is the basic idea of the resolution. The discussions of the previous Section \ref{sec:cr1} take a notion of isomorphism for granted and give an account of these isomorphisms through the automorphisms of a structured base space, by satisfying Desideratum (i). Thus, I argued, invariant structure could be understood qualitatively.  Now we will  recover, or build-up, the space of models and their isomorphisms by first accepting certain physically meaningful  patterns as the basic constituents of a theory. From a metaphysical standpoint, we would like to know what there is, first, and then deduce from that physical input some representation through models and  a notion of model-isomorphism.  %Accepting \cite{Moller}'s Motivationalism, we should feel motivated to  verify that the theory can be given such an account, satisfying  (ii). 

%But this resolution will \emph{not} answer  Challenge (2), of expressing every individualistic fact using qualitative facts.

The axiomatic approaches to be described below---which are not new or my own---are labeled \emph{synthetic}.  Their basic axioms  carry elementary \emph{physical} meaning in the theory.\footnote{In \cite{AdlamLinnemannRead} such approaches are labeled `constructive', but  for philosophers that term strongly connotes  intuitionism, \emph{a l\'a} Brouwer and cousins. According to \cite{AdlamLinnemannRead}, such approaches should be distinguished from the mere \emph{deductive approaches}, that find uniqueness results from more basic postulates that don't necessarily carry a physical interpretation. } This feature is what allows us to intepret the resulting mathematical structure physically, and thus provide a metaphysically perspicuous picture of the underlying ontology. 
 
 In Section \ref{sec:GR_case} and in Section \ref{sec:YM_case} we will discuss synthetic axiomatizations of spacetime metrics and  connections in principal fiber bundles,  respectively.  

 \subsubsection{The case of spacetime metrics.}\label{sec:GR_case}
Here I will focus on two related approaches to axiomatization:  \citep{EPS} and \citep{Mundy1992}.\footnote{\citet[p.518]{Mundy1992} admits  kinship with \cite{EPS}, and even describes the latter as ``deeper'', but also condemns their argument as ``sketchy and incomplete''. I do not aim to assess these matters: this Section is illustrative only. For a complete appraisal of EPS and kindred synthetic approaches, see \cite{AdlamLinnemannRead}. }  

 We start with  the seminal \cite{EPS}, who set out to successively build differential topology, conformal, affine, and Lorentzian structure, from basic postulates concerning the paths of light and massive particles (see also \cite{LinnemannRead} for a pedagogical introduction to this scheme of Ehlers, Pirani and Schild (henceforth EPS)).  
  According to \cite[p.2]{LinnemannRead}, the EPS scheme is ``constructively axiomatic in the sense of Reichenbach---
i.e., builds on a basis of empirically supposedly indubitable posits.'' The two main initial posits define: (1)  A point set $M = \{p,q,...\}$, called a set of events; and (2) 
 The elements of $L = \{L,N,...\}$ and $P =
\{P,Q,...\}$, where $L,N ..., P,Q,...$ are all subsets of $M$, and are respectively called `light
rays' and `particles'. Of course,  it is too early to call $L$ and $P$ ‘light rays’ and ‘particles’: they will only acquire this meaning once they are subject to further empirically-motivated posits, such as the assumption that one can bounce a light ray from the worldline
of one particle to that of another and back again, etc. Indeed, $(M, P,L)$ only acquire their standard interpretation at the end of the construction,  once they have been employed to define increasing levels of structure (smooth, affine, projective, etc).

 But let me focus on this Section on  \cite{Mundy1992}, who also seeks an intrinsic geometric axiomatization of semi-Riemannian geometry.
  Such an axiomatization follows in the tradition of many others (e.g. \cite{Robb1936}, for the case of Minkowski space). Thus, \citet[p. 517]{Mundy1992} uses the intrinsic geometric axiomatization of Euclidean geometry as a template. There, we use coordinate-free primitives such as the \emph{affine betweenness} relation $B(p,q,r)$, read as `$q$ is between $p$ and $r$ on a straight line-segment', the \emph{segment congruence} relation $C(p,q,r,s)$, read as `the straight line-segments between $p$ and $q$ and between $r$ and $s$ are  congruent', etc.  to formulate a theory $T_E$ in a language $L_E$ including these primitives relations. Equivalence with the coordinate formulations is shown by a representation theorem: that each model of $T_E$ has Cartesian coordinates, unique up to orthogonal transformations, representing $B$ and $C$ by the standard coordinate formulas.

Analogously, Mundy's  axiomatization of (semi-)Riemannian geometry proceeds in terms of  ``truly coordinate-free primitives'' in which $C$ refers to metric behavior of clocks and rods, and $B$ refers to geodesic motion. In analogy to the Euclidean case, one replaces straight lines by geodesics, and has $B(p,q,r)$ as the relation of \emph{geodesic betweeness} and  $C(p,q,r,s)$ is a notion of \emph{geodesic congruence}. Analogously to  the Euclidean case, a third intrinsic relation,  $A(p,q,r)$, of \emph{geodesic orthogonality}, is defined infinitesimally in terms of $B$ and $C$, by using infinitesimal trigonometry. The same can be done with $S(p,q,r,p'q')$, the notion of \emph{projective separation} (which I will not recount here). The general idea is that (semi)-Riemannian geometry can, as regards its local structure, be developed in a synthetic manner, along similar lines to Euclidean geometry.

The upshot  is that this language, $L_R$
has no sentence which would be made true by a model  $S:=\langle M, g\rangle$, but not by $Sf:=\langle M, f^*g\rangle$ (or vice versa).  
 In the words of \citet[p. 520]{Mundy1992}:
\begin{quote} the difference between the two models [$S$ and $Sf$] is not expressible in the theoretical language $L_R$: they are isomorphic, and therefore satisfy all of the same sentences. (Note that this is theoretical, not merely observational, equivalence.) [...] This is because the language $L_R$ does not and cannot contain any term which ``rigidly designates"  the point p itself [i.e. is non-qualitative], i.e., which refers to p when $L_R$ is interpreted over $S$, and also refers to p when $L_R$ is interpreted over $Sf$. \end{quote} 

Thus,  while the metalanguage is able to express singular, non-qualitative features, $L_R$ is not. Isomorphisms, expressed in the metalanguage, will map objects singled out by the same description into each other.
Thus Mundy's  formal semantics under some interpretation defines a structure  admitting a notion of isomorphism that coincides with isometry.

I take the   synthetic approach to spacetime geometry illustrated here  (twice) to provide a structure-first definition of the theory; and I also take it  to answer the  metaphysician's misgivings above, about a qualitative description of spacetime structure. 

To nail the point home, take \citet[p. 147]{Dasgupta_bare}'s  first challenge as representative of that concern.\footnote{As quoted at the start of Section \ref{subsec:skeptic_why}, the challenges are: ``(1) to clearly articulate what the underlying qualitative facts are like, and (2) show that they are sufficient to explain (in the metaphysical sense)
individualistic facts about the manifold.} He proposes (p. 149, ibid) that a solution would need a language ``PL of predicate logic [...] in which every predicate expresses a qualitative property or relation.  Every fact that can be expressed in PL is a purely qualitative fact.''  %So this language is suited to expressing certain facts, such as  that expressed with the 2-place sentence ``someone loves someone'', or $(\exists x, y) (x$ loves $y)$. 
%(The idea is fundamentally \emph{holistic} (p. 151, ibid): each world is described all at once by a complete sentence in this language, with a possibly infinite number of connectives.)

  Both \citep{Mundy1992} and \citep{EPS}  furnish us with  this kind of language: certain relations between events, such as `being (non)causally connected', or `having the same length', or even `being connected by a massive particle's worldline', are primitive and qualitative, just like `having non-zero absolute velocity' is a qualitative property in Newtonian mechanics (cf. footnote \ref{ftnt:abs_vel}); or just like the sentence ``someone loves someone'' in predicate logic.

As to a metaphysically perspicuous interpretation, the constraints on these languages allow solely for the expression of isomorphism-invariant (i.e. diffeomorphism-invariant) quantities, whose  interpretations are by construction attached to primitive physical posits, about point-coincidences, proper times along worldlines, etc. 

In sum, jointly, we  physically interpret general relativity's fulfillment of Desiderata (i) and (ii) as  taking diffeomorphism-invariant mathematical quantities to represent physically significant quantities, understood as quantities about coincidences of material point-particles, elapsed proper times along a particle worldline, etc. Though  our models of these physical features are defined only up to isometry, this poses no threat to determinism: we can identify physical possibilities   as given by an invariant structure which can be interpreted   anti-haecceitistiscally.%; which can be interpreted as mere changes of coordinates describing \cite[Sec. 5.3]{Samediff_0}.  %For example, if one makes a journey from one planet to another, all empirically measurable quantities  about the trip will be represented as diffeomorphism-invariant functions.% These include: the time elapsed along the journey, whether the spaceship is intrinsically. 

 \subsubsection{The case of connections on a principal bundle.}\label{sec:YM_case}
 Similarly, we have two examples of axiomatization for Yang-Mills theory. These are at the very least as intricate as the previous case of general relativity. But for reasons of space, I will only briefly gloss the two constructions here. 
 
 The first is based on \cite{Barrett_hol}'s construction of a general connection $\omega$ on a principal fiber bundle, $P$, as discussed in \cite[Sec. 4]{Samediff_0}. Here, besides the spacetime structure, which could be constructed as in the previous Section, we can take the basic axioms to have physical meaning. The physical meaning of these postulates follow the prescription sketched in \cite{YangMills}'s quote at the end of \cite[Sec. 4]{Samediff_0}, and argued at length in that Section. That is, the primitive, empirically-motivated axioms, refer to the relative rotation of a charge, obtained by dragging it (i.e. parallel transporting it) along  closed curves of the spacetime manifold and comparing the result with the initial charge.\footnote{In practice, this would be accomplished with both charges undergoing time-like motion, with their joint trajectories forming a closed loop.} In Yang-Mills theory, this overall rotation is called \emph{the holonomy} (we will discuss it in more detail in \cite{Samediff_2}; see \cite{Healey_book} for a philosophical introduction).  From these basic quantities, we can construct a principal fiber bundle and, on it,  a unique Ehresmann connection-form $\omega$ (cf. \cite[Sec. 4]{Samediff_0}), up to vertical automorphisms (the isomorphisms of the principal fiber bundle structure).
 
 In more detail, let $M$ be the spacetime manifold, $G$ be a Lie group, and $\gamma$ be a  loop based at point $x$, namely $\gamma: [0,1]\rightarrow M$, with $\gamma(0)=\gamma(1)=x$. The space of all such loops we call $\text{Loop}\,(M)$, and we postulate a map:
 \be hol: \text{Loop}\,(M)\rightarrow G,
 \ee
 which we take as primitive, and obeying three postulates, which can all be empirically motivated (see \cite[Sec. 2.4]{Barrett_hol}):
 \begin{enumerate}[(H1)]
\item  $hol$ is a homomorphism of the composition law of loops and group multiplication:
$$ hol(\gamma_1\circ \gamma_2)=hol(\gamma_1)hol(\gamma_2).$$
\item 
$hol$ takes the same value on  loops which
differ by a reparameterisation, or by the addition or
removal of path sections which `double back' on
themselves. Thus, if $\gamma_1\circ\gamma_2$ encloses zero area, 
$$hol(\gamma_1\circ\gamma_2)=\mathrm{Id}.$$
Both these postulates have a straightforward empirical interpretation, used in the construction theorem to interpret $hol$ as the parallel transport operation for internal vectors. 
\item 
The last condition requires topological structures in the space of loops; general constraints select a unique family of topologies. Given one such  specific topology and a smooth family of loops, say $\ell: \RR\rightarrow \text{Loop}\,(M)$ (we could replace $\RR$ for any other parametrizing space, e.g. $\RR^2$, for a 2-parameter smooth family of loops), then:
$$  hol\circ \ell \quad\text{is also smooth}.
$$ 
Barrett interprets this as a  condition on physical approximations: if we were to physically vary the loop corresponding to a holonomy slightly, the end result of the holonomy must also be varied only slightly. 
 \end{enumerate}
 With these postulates, Barrett proves a \emph{construction} and a \emph{presentation} theorem,\footnote{In fact, he calls them \emph{re}construction and \emph{re}presentation, but these are used in the opposite sense of \cite{Mundy1992}: for instance, his reconstruction goes from the physically transparent synthetic theory to various representations in bundles.} 
 showing, respectively that: (1) for $M$  a connected
manifold with basepoint $x$, and $hol: \text{Loop}\,(M)\rightarrow G$ satisfying 
conditions H1-H3,  then there exists a
differentiable $G$-principal fibre bundle P, a point $p\in \pi^{-1}(x)$ and a connection $\omega$ on $P$ such that
$hol$ is the holonomy mapping of  this bundle and connection; (2) the presentation theorem shows that for any $(P, \omega)$ such that $\pi:P\rightarrow M$, there is a holonomy mapping on $M$ that recovers $(P, \omega)$ up to a vertical isomorphism.
 And indeed, \cite{Barrett_hol} also construes his theorems as empirically motivated.\footnote{And the empirical motivation also applies for the reconstruction of the spacetime connection through parallel transport along loops (or holonomies): 
 \begin{quote}
 The information in the gravitational field is involved
in grouping together, in a particular way which will be
described more fully below, all the possible particle
paths into sets which represent the particles whose paths
are "in coincidence". These sets form the points of the
space-time manifold. This is actually all that is
required! What one finds is that the geometry of the
particle paths is sufficient to specify the geometry of
the resulting space-time. The result is a theory of
gravity in which the test particles are firmly mixed up
with the phenomenon of the field itself. In fact the field
is "made out of" the motions of the test bodies. There is
no conception of an independent gravitational field
divorced from the theory of the propagation of matter. (p.22, ibid)\end{quote}
}

  The second sort of axiomatization---which we will here only very superficially gloss---is based on the joint axiomatization of spacetime and of a Lie-algebra; and the latter proceeds in a similar fashion to the axiomatization of any finite-dimensional vector space, interpreted as a space of internal  vectors (e.g. colours or charges). That is, the Yang-Mills connection is an object that mixes tensorial indices with internal indices. The natural principal bundle for the tensorial part,  (see \cite[Sec. 3]{Weatherall2016_YMGR} and \cite[Sec. 4]{Samediff_0}), would be a sub-bundle of the  frame bundle $L(TM)$; and the internal part to the vector bundle associated to $\mathfrak{g}$.
These two bundles are of different types, and to make a joint object out of them, we need  `bundle splicing'  (see e.g. \cite[Ch. 7.1]{Bleecker}).
 The structure that emerges from this splicing is known as the `bundle of connections', or the Atyiah-Lie algebroid, and it represents sections of $TP/G$, which are in 1-1 correspondence with Ehresmann connections, $\omega$. See \cite[Ch. 17.4]{Kolar_book} for the construction alluded to here, which is discussed in  further detail in \cite[Sec. 6]{Samediff_0} and employed in \cite{Samediff_2}. \\

In sum, the moral, for the two types of theory, is the same: The isomorphisms that appear in the higher level language  can  be understood through a  representation theorem for models of the more basic ontology of the theory. Once we have the Ehresmann connection $\omega$, and a spacetime metric $g_{ab}$, built from more fundamental, empirically motivated posits, we can stop appealing to this more cumbersome axiomatic language and proceed in the standard textbook (metalanguage) terms; thereby we recover the  isomorphisms of the theories   (used in Assumptions II and III of the introduction to Section \ref{sec:responses}). 

The idea of the extension of sophistication to more general theories then, is that, were they to satisfy Desiderata (i) and (ii),  they would enjoy the same kind of metaphysically perspicuous interpretation of the symmetry-invariant ontology, and that anti-haecceitism/anti-quidditism would suffice to physically identify symmetry-related models.  
 
 %Thus,  it is natural to articulate the ontic commitments of gauge theory that ensue  as follows:  each possible world---each particular choice of structure---is given by  one way in which internal quantities are parallel transported over spacetime. In a particular formulation of general relativity that is suited to the analogy: each possible world is given by one way in which external quantities are transported over spacetime.%Or, more explicitly: given by how the various  sorts of charges vis a vis some force or interaction---interactions that can be classified by Lie algebras---are to be identified along spacetime directions. Indeed, in this structural interpretation,  the Aharonov-Bohm effect (cf \cite[Section  2]{Samediff_2}), is not mysterious at all: it merely manifests global facts about parallel transport. 
%And the structural  interpretation of Yang-Mills theory is conceptually similar to the  structural interpretation of general relativity. For general relativity, each possible world, or physical possibility---each particular choice of structure---is given by one way in which spacetime points are chronogeometrically related or distributed. For Yang-Mills, the structure  represented by $\omega$ refers to the parallel transport of internal quantities.

 %, and these can be understood `internally', in the sense of \cite{Dewar2017}. 

\section{Summing up }\label{sec:conclusions_soph}
%\section{Anti-haecceitism and anti-quidditism: sauce for the gander, but not for the goose? }\label{sec:conclusions}

%Most theories are given as in (i), but, accepting \cite{Moller}'s Motivationalism, we should feel motivated to then verify that the theory can be given an account satisfying  (ii). 

This paper has developed a view of sophistication that could be extended to more general theories and that applies to both spacetime diffeomorphisms and gauge symmetries.

The view was employed to defend the doctrine of anti-quidditism for gauge theory, but, more emphatically, the doctrine of anti-haecceitism for spacetime. For anti-haecceitism is still actively disputed today: in essence, the concern is that it could be a mere stipulation that does not bear the metaphysical interpretation foisted upon it; for the doctrine simultaneously deflates the individuation of spacetime points and reaffirms the existence of spacetime. That concern, I argued, has a close cousin in the modern literature about sophistication: the concern that the ontology that is common to all of the symmetry-related models is not sufficiently explicated by stating that symmetry-related possibilities are to be identified---\emph{that} is also mere stipulation!

My Desideratum (i) for sophistication guaranteed that physical indeterminism could be eliminated by anti-haecceitism, or, more generally, by commitment to a purely qualitative ontology. By satisfying (i), a theory's isomorphisms emerge from the automorphisms of the base set of the theory's models. Thus isomorphic models differ only as to \emph{which part} of the base set supports \emph{which pattern}: isomorphic distributions are `qualitatively' identical, in a precise sense of qualitative. By requiring the isomorphisms to be related to the dynamical symmetries of the theory and to the automorphisms of the base set, Desideratum (i) constrains, and may suffice to entirely regiment, the idea of `internal' sophistication. And it guarantees that there are no empirical differences between isomorphic models. But this Desideratum takes the space of models $\F$ for granted, and with it a notion of isomorphism of the models. Desideratum (i) does not \emph{start from} the invariant structure---it does not, strictly speaking, take the `structure first'---and it falls short of ensuring a `metaphysically perspicuous' construal of the structure that is common to the isomorphism classes. 

Fortunately, Desideratum (ii) for sophistication picks up where Desideratum (i) stops.  According to Desideratum (ii) we should understand the invariant structure first, `internally': or rather, synthetically. This is accomplished explicitly by requiring the invariant structure to be defined axiomatically. The isomorphisms then relate different representative models of the same structure. Moreover, the axioms explicitly endow the invariant structure with a perspicuous (meta)physical interpretation, since they are expressed using basic physical posits of the given theory. In this way, we conceptually clarify the isomorphism-invariant mathematical structure while connecting that structure to the basic ontology of the theory. 

Were a theory to fill Desideratum (ii) but not Desideratum (i), we would not be able to interpret the invariant structure `qualitatively' (at least in a well-regimented sense). And we would not be able to say whether an ontological commitment to the invariant structure would suffice to eliminate the threat of physical indeterminism.

 I take this to close the question of whether we should endorse sophistication for spacetime diffeomorphisms, but not for gauge symmetries; that is, whether we should, in these cases, be (resp.) anti-haecceitists but not anti-quidditists. In both cases, and others that fulfill Desiderata (i) and (ii) as well, we should be `anti-ists'.\\

 I will end this paper with a view towards the next in the series, \cite{Samediff_1b},  by turning the gaze of our investigation `inward', that is, to objects \emph{within} each model. While  sophistication as construed here also  implies  objects within each model can be  individuated only by the qualitative properties which the models tack onto them, the doctrine is silent about what is required for uniquely specifying objects inside each physical possibility. This is related to the second worry, about structure-tokens, glimpsed in Section \ref{sec:method}, and it  will be the topic of the third paper \cite{Samediff_1b}.

\subsection*{Acknowledgements}

I would like to thank: first and foremost Jeremy Butterfield for many conversations on this topic, and for reading several versions of the paper;  Ruward Mulder, for a careful reading and comments; and Caspar Jacobs, for reading and providing encouragement and suggestions. %And I would also like to thank two anonymous referees, for their patience and   comments, which led to several revisions of an early draft. 

%Of course I could not conclusively show that the structural representation is equally recommended in both cases; I can only show that it is equally conceptually transparent in both cases. In the following Section I  will assess  some authors' concrete attempts at drawing distinctions between general relativity and gauge theory, seeking to strengthen \emph{the license} for sophistication of gauge symmetries to \emph{a recommendation}. So my final advocacy for this recommendation will be completed in the accompanying paper \cite{Samediff_2}.    

%This finishes this Section's analysis of the conceptual, high-altitude similarities between gauge symmetry and diffeomorphisms. Now  I want to move on to a discussion of their differences. %, and for that we can focus once again entirely on the \emph{active} symmetries of the theory (these transformations are easier to manipulate at an abstract level, and they suffice for our proposed  distinction between gauge from diffeomorphism symmetry. 
% In short, we are now ready to evaluate the more concrete similarities and differences between the two theories, concerning especially their symmetries. 
 % \bibliographystyle{apacite} 
% \bibliographystyle{alpha}
%\bibliography{references2}

 \end{document}